\documentstyle[aps,manuscript]{revtex}
\begin{document}
\draft
\title{Parallel updating cellular automaton models of
driven diffusive Frenkel-Kontorova-type systems}

\author{B.H. Wang$^{1,2,3}$, Y.R. Kwong$^{4}$, P. M. Hui$^{4}$,
and Bambi Hu$^{1,5}$}

\address{$^{1}$Department of Physics and Centre for Nonlinear Studies,\\
Hong Kong Baptist University, Hong Kong, China\\
$^{2}$Department of Modern Physics and Nonlinear Science Center,\\
University of Science and Technology of China, Hefei 230026, Auhui, China\\
$^{3}$CCAST (World Laboratory), P.O. Box 8730, Beijing 100090, China \\
$^{4}$Department of Physics, The Chinese University of Hong Kong,\\
Shatin, New Territories, Hong Kong, China \\
$^{5}$ Department of Physics, University of Houston, Houston, TX 77204}

\maketitle

\begin{abstract}
Three cellular automaton (CA) models of increasing complexity are 
introduced to model driven diffusive systems related to the generalized 
Frenkel-Kontorova (FK) models recently proposed by Braun {\em et al.} 
[Phys. Rev. E {\bf 58}, 1311 (1998)].  
The models are defined in terms of parallel updating rules. 
Simulation results are presented 
for these models.  The features are qualitatively similar to those 
models defined previously in terms of sequentially updating rules.  
Essential features of the FK model such as phase transitions, jamming 
due to atoms in the immobile state, and hysteresis in the relationship
between the fraction of atoms in the running state and the bias field 
are captured.  Formulating in terms of parallel updating rules has 
the advantage that the models can be treated analytically by following 
the time evolution of the occupation on every site of the lattice.  
Results of this analytical approach are given for the two simpler models. 
The steady state properties are found by studying the stable fixed 
points of a closed set of dynamical equations obtained within the 
approximation of retaining spatial correlations only upto two 
nearest neighboring sites.  Results are found to be in good 
agreement with numerical data.

\end{abstract}


\newpage

\section{Introduction}

The physics of driven diffusive systems has attracted much attention
recently \cite{privman,zia} due to their relevance to the general area of
nonequilibrium statistical mechanics and their wide range of possible
potential applications.  In particular, the Frenkel-Kontorova (FK)
model \cite{fk,2dcrystals} and its 
generalizations\cite{perrson,elmer,rozman}
have been studied within
the context of tribophysics.  Braun and coworkers \cite{braun1,paliy,braun2} 
studied
the atomic current in one and two-dimensional atomic systems in the 
presence of a
periodic potential under the influence of a dc driving force within
the approach of Langevin equations.  In tribophysics, 
the driving force emerges owing to
the motion of one of the two substrates separarted by a thin atomic
layer.  The results of these studies are characterized by two features.
One feature is that the system exhibits hysteresis in response to the
driving force.  The system jumps between low-mobility and high-mobility
regimes in a hysteretic manner as a function of the driving force.  Another
feature is that accompanying this transition, the atoms tend to organize
themselves into two types of domains consisting of atoms in states
of different characters, one consisting of slowly moving (``immobile") atoms
and another consisting of ``running" atoms moving with maximum speed.
The latter feature resembles those in traffic flow models\cite{ns,fi,chung}
in which cars
may be moving at its maximum speed if they are not blocked or may be
momentarily stationary if they are blocked in front.  

The models studied in Ref.\cite{braun1,paliy,braun2} 
are quite complicated.  Attempts have
been made \cite{braun3} to introduce simpler models which capture the
essential features.  In a series of three lattice gas (LG)
models of increasing
complexity (henceforth referred to as LG Models A, B, and C),
Braun {\em et al.} introduced probabilistic hoppings of atoms on a lattice
together with the possibility of the atoms being found in one of two
possible states.  The underlying model (LG Model A) in one dimension
is that $N$ atoms are placed in a lattice of length $L$ corresponding
to a concentration of $\rho = N/L$.  The dynamics is introduced in a
random and sequential fashion by randomly choosing a site at each time step
and updating the system according to specific rules.  The hopping to
the nearest neighbouring sites of an immobile atom in a randomly chosen site
is characterized by a probability $\alpha$  ~($1-\alpha$) that an atom hops
into the site in the right (left) hand side.  Thus the parameter
$\alpha$ models the effect of a driving force, and the hoppings to
the right and left hand sides correspond to the effects of drift together
with diffusion.   Atoms in the running state always attempt to hop to the
right, which   
is taken to be the direction of the driving force.  
The $\alpha = 1/2$ case corresponds to vanishing bias field.   
The $\alpha=1$ case corresponds to the totally asymmetric exclusion
model, which has been solved exactly \cite{derrida}.  An atom in the immobile
state changes to the running state if it succeeds in hopping to the right
hand side, while a running atom becomes immobile if the site to the right
is occupied by an atom in the immobile state.  Such transitions between
the immobile and running states of an atom model the effects of damping.
LG Model A is thus characterized by the parameters $\rho$ and $\alpha$.
LG Model B includes the possibility that the running atom at a randomly
chosen site may change to the immobile state with probability $\gamma$
prior to the motion of the atom takes place.  This spontaneous convection
from the running to the immobile state is supposed to be more important for
weak driving forces.   In order to capture the 
features of hysteresis, Braun {\em et al.}\cite{braun3} 
went on to include in the basic model
a majority rule in the conversion between the two states of the atoms.  If
the randomly chosen site is occupied by the leading atom in a compact
group of $r$ running atoms and it is blocked by a compact group of $s$
immobile atoms in front, all the $r+s$ atoms will turn into the
immobile (running) state if $r <s$  ($r \geq s$).  These three models have
the advantage of being easily implemented numerically by carrying out
Monte Carlo simulations.

It is useful to study similar models within the context of dynamical systems.
A physical consideration is that parallel updatings of the states of
the atoms may be more appropriate than the sequential updatings in the LG
models studied in Ref.\cite{braun3}.  In the present work, we propose
analogous models with parallel updating rules 
in that all the atoms evolve in every time step according
to updating rules, and hence the models become cellular automaton (CA).
We performed numerical simulations on the models.  Another advantage of
casting the models in terms of 
parallel updating rules is that a more systematic
analytical approach, analogous to those successfully applied to
traffic flow models\cite{ss,wang}, may be applicable.  Such approach focuses
on the time evolution on the state of each of the sites, namely whether
the site is occupied by an atom in the running or immobile state or
unoccupied.  In general, spatial correlations of gradually increasing
spatial extent are introduced as time evolves.  Equations can be written down
relating the state of a site at time $t+1$ to quantities at time $t$.
By suitably decoupling the spatial correlations, a set of coupled nonlinear
equations can be obtained with the fixed point corresponding 
to the solution
in the long time limit.  The complexity of the set of equations depends
on the extent of spatial correlations retained after decoupling.  
The approach has the advantage that it gives the fraction of running 
atoms in the steady state together with other spatial correlation 
functions. 
As an
illustration of the general idea of the approach, we study Models A and B
with parallel updating rules and results are found to be in reasonable
agreement with numerical simulations within the approximation of retaining
correlations upto two sites.  The present work, therefore, complements 
that of Braun {\em et al.} \cite{braun3} and suggests an alternative 
way of studying the various models proposed within the context of 
tribology.  

The plan of the paper is as follows.  In Sec.II, we define the modified
models with parallel updating rules and present the results obtained by 
numerical simulations.  Section III reports results of our analytical
calculations on model A and B.  Results are compared with numerical
simulations.  We summarize the results in Sec.IV.

\section{Models and numerical results}

\subsection{Model A}
We consider $N$ atoms on a one-dimensional lattice of $L$ sites
with periodic boundary condition.  Following the 
basic lattice gas model\cite{braun3}, we
modify the rules such that parallel updating is incorporated.  The updating
rules are:

\vspace*{0.1 true in}
\noindent {\em Rule A1}:  If the $i$th site on the lattice at time $t$
is occupied by an immobile atom, it has a probability $\alpha$ to be in
the advancing immobile state, i.e., state that flavours forward biasing 
 and a 
probability $1-\alpha$ to be in the retreating immobile state,
i.e., state that
flavours backward hopping 
.  An advancing immobile atom can either hop into the
{\it empty} ($i+1$)th site {\it and become a running atom} if the ($i+2$)th 
site is {\it not} occupied by a retreating atom or be blocked by an 
atom occupying the 
($i+1$)th site without changing its state, while a retreating 
atom can either hop into the {\it empty} ($i-1$)th site {\it and stay in  
the immobile state} if the ($i-2$)th site is {\it not} occupied by a running 
or an advancing immobile atom or be blocked by an atom occupying the
($i-1$)th site without changing its state. 

\vspace*{0.1 true in}
\noindent{\em Rule A2}: If the $i$th site on the lattice at time $t$ is
occupied by a running atom, it can 
move to the {\it empty} ($i+1$)th site if the ($i+2$)th site is {\it not} 
occupied by a retreating atom or stay at the $i$th site
and remain in the running state if it is blocked by a running atom
or {\it change to the immobile state} if it is blocked by an immobile atom 
at the ($i+1$)th site. 

\vspace*{0.1 true in}
\noindent{\em Rule A3}: If the $i$th site on the lattice at time $t$ is 
empty and is sandwiched between
a running or advancing immobile atom at the ($i-1$)th site and 
a retreating immobile atom at the ($i+1$)th site, then the atoms at the two
neighboring 
sites are equally probable to hop into the $i$th site according to 
rules A1 and A2.  The atom that fails to hop at that time step will remain 
in its original state.

Model A thus represents a modification of the LG Model A in Ref.\cite{braun3}
with parallel updating rules.  The quantity of interest is
the mobility $B$ of the system which is defined as the fraction of
atoms in the running state in the long time limit.  We have carried out
numerical simulations on Model A.  Figure 1 shows the dependence 
of the
mobility $B$ as a function of the drift parameter $\alpha$ for
different concentrations of atoms $\rho$.  
Note that only the range $1/2 < \alpha \leq 1$ corresponding to a biased
field to the right hand side is shown, although the range $0 < \alpha<1/2$ 
can also be studied taking Model A as a CA model in its own right.  To 
illustrate the basic features of the model, we have performed numerical 
simulations on systems with $L=1000$.  Typically, about 2000 time steps 
are sufficient for approaching the long time limit.  The mobility $B$ is 
obtained simply by counting the fraction of running atoms in the long
time limit.  For $\alpha <1$, an average over 50 random initial 
configurations are performed.  For $\alpha > 1/2$, $B=1$  for $\rho <0.32$. 
For $0.32 < \rho <1/2$, the mobility $B$ becomes unity at some critical 
value $\alpha_{c}(\rho)$.  For $\rho > 1/2$, the concentration is 
sufficiently high so 
that $B<1$ for all values of $\alpha$.  

The particular point of $\alpha=1$
deserves further discussion.  It is found that for $\rho > 1/2$, the 
mobility $B$ at $\alpha=1$ in the long time limit depends on the initial condition. 
The results at $\alpha =1$ shown in Fig.1 correspond to $B = 
(1-\rho)/\rho$, which are obtained by using the initial configuration 
in which all the atoms are immobile.  For arbitrary initial 
configurations at $\alpha =1$, $B$ is found to lie within the 
range $(1-\rho)/\rho \leq B \leq 1$.  Similar results are obtained 
in our analytical approach, discussed in the next section, by treating the 
model as a dynamical system.

\subsection{Model B}
Model A forms the basic CA model for further modifications.  In particular,
while irreversible transition into the running state for an isolated
atom is strongly favourable in the high-field limit ($\alpha \approx 1$),
it is possible for a running atom to convert spontaneously to the immobile
state in the weak field case.
Following the LG model B in Ref.\cite{braun3}, we introduce the following 
rule in addition to the rules A1, A2, and A3 stated above:

\vspace*{0.1 true in}
\noindent {\em Rule B1}: Before the updating rules A1, A2, A3 are applied in each time
step, every atom in the running state has a probability $\gamma$ to change
its state to the immobile state and a probability $1-\gamma$ to remain
in the running state.  After this consideration, all the atoms on the lattice
evolve according to the rules A1, A2, and A3.

Rules B1, A1, A2, and A3 define the CA Model B.  
While $\alpha$ tends to lead to a larger fraction of running atom, the
parameter $\gamma$ counteracts the effect and tends to increase the number
of atoms in the immobile state.  Hence, the mobility $B$ is generally lower
for $\gamma \neq 0$ cases than the $\gamma = 0$ case
for the same value of $\alpha$.  Figure 2 shows the values of $B$ as a function of
$\alpha$ for different values of $\gamma$ with the concentration fixed at  
$\rho=0.4$, which corresponds to a concentration at which the atoms are 
isolated if they are uniformly distributed on the lattice.  Note that the 
mobilities
converge to unity at $\alpha=1$ for different values of $\gamma$.   
Only when $\gamma=0$ will the mobility becomes unity for $\alpha_{c} < 
\alpha \leq 1$.  The results shown are typical for $\rho < 1/2$.  Figure 3
shows the results for $\rho=0.6$, which corresponds to a concentration at 
which there are always some atoms with 
nearest neighbors if they are uniformly 
distributed on the lattice.  In this case, $B<1$ for all values of $1/2 
< \alpha <1$ and $0\leq\gamma \leq 1$.  The mobilities 
converge to the same value for different values of $\gamma$ at 
$\alpha=1$.  For $\gamma = 0$, Model B reduces to Model A and 
the mobility $B$ lies in 
the range $(1-\rho)/\rho \leq B \leq 1$ with the precise value depending 
on the initial condition.

\subsection{Model C}
Following the LG model C in Ref.\cite{braun3}, further modifications  
can be made by taking
into account the influence of the state of the 
surrounding atoms on that of 
a single atom, i.e., the `crowding  
effect'.  The modifications involve
the considerations of the jamming of a running block of atoms (i.e., a 
compact group of nearest neighboring atoms in the running state) by    
an immobile block of atoms (i.e., a compact group of atoms in the immobile 
state).  If these two adjacent blocks of atoms are 
sandwiched between two empty sites at the two ends, 
the state of {\it all} the atoms will then follow that of the 
larger block.  This rule will be referred to as the majority rule.  
Furthermore, 
since the spontaneous transition of atoms in the
running state to the immobile state should be suppressed by an  
increasingly stronger forward 
bias, 
the parameter $\gamma$ should be $\alpha$-dependent.  
We impose the relation \cite{braun3}
$\gamma = \gamma_{0} (1-\alpha)^{2}$ on the parameter $\gamma$,  
where $\gamma_{0}$ is a model parameter.  Hence, the
updating rules for Model C can be state explicitly as below: 

\vspace*{0.1 true in}
\noindent {\em Rule C1}: At a certain time step, if the ($i-r+1$)th site to 
the $i$th 
site are {\it all} occupied by running atoms and the ($i+1$)th site to the 
($i+s$)th site are {\it all} occupied by immobile atoms together with the 
condition that the ($i-r$)th site and
the ($i+s+1$)th site are empty, then in the case of $r\geq s$ ($r<s$), 
{\it all} 
the $r+s$
atoms become running (immobile).  Immediately after the changes, the 
states of the sites are then updated according 
to the rules of Model B with $\gamma =\gamma_{0} (1-\alpha)^{2}$.

Rule C1 together with Model B define the CA Model C.  
Figure 4 shows typical results for 
the mobility $B$ as a function of $\alpha$ for different
values of
$\rho$ with $\gamma_{0}$ taking on a value close to $0$.  It is observed 
that the results look 
very similar to that of Model A except for the presence of {\it 
hysteresis}, a characteristic feature of the FK model \cite{braun3}, for
$\rho \leq 1/2$.  The values of $B$ obtained by 
gradually increasing $\alpha$ from $\alpha=0.5$ to
$\alpha=1$ are generally smaller than those obtained by gradually 
decreasing $\alpha$ from $\alpha=1$, and the difference between the 
mobilities for increasing and decreasing bias fields at a particular 
value of $\alpha$   
increases with $\alpha$ as observed in the LG model in Ref.\cite{braun3}.

\section{Analytical approach}
The CA models with parallel updating rules have the
advantage that they can be treated analytically
within the context of dynamical systems.  The general idea is to establish
the time evolution equations for the state on each site of the lattice.
The equations, in general, involve spatial correlation functions.  With
suitable approximations typically involving proper decoupling of
the correlations, a closed set of dynamical equations can be obtained.  Such
a set of equations can be treated as a dynamical mapping between quantities
at time $t+1$ and those at time $t$.  Hence, following standard approaches
in dynamical systems, the solution in the long time limit can be
found by studying the fixed points (attractors) of the set of equations.
Such an approach has been successfully developed for traffic flow
models \cite{ss,wang} in which the cars,
which is analogous to the atoms in the present
models, can only move in one direction without backward diffusion movements.
To illustrate the idea, we apply the approach to study the modified
CA models.  It turns out that for Models A and B, reasonably good
agreement with numerical simulations can be obtained by retaining
spatial correlations involving two neighbouring sites only.  Application
of the method to Model C is difficult due to the built-in
long spatial correlations in the majority rule of the model, and hence 
results are only reported for Models A and B.  

\subsection{Model A}
Although one can treat Model B directly and obtain results of Model A by  
setting $\gamma=0$, it is, however, illustrative to treat the simpler 
Model A first. 
We denote the states of the $i$th site at time $t$ 
by the following set of 
boolean variables:
\begin{eqnarray}
R_{i}(t)&=&\left\{
	\begin{array}{ll}
	1, &{\rm if\,the\,}i\,{\rm th\,site\,at\,time\,}t\,{\rm is\,occupied\,
by\,a\,running\,atom} \\
	0, & {\rm otherwise}
	\end{array}
	\right.  \\
I_{i}(t)&=&\left\{
	\begin{array}{ll}
	1, &{\rm if\,the\,}i\,{\rm th\,site\,at\,time\,}t\,{\rm is\,occupied\,
by\,an\,immobile\,atom} \\
	0, & {\rm otherwise}
	\end{array}
	\right.   \\
S_{i}(t)&\equiv&R_{i}(t)+I_{i}(t) 
=\left\{
	\begin{array}{ll}
	1, &{\rm if\,the\,}i\,{\rm th\,site\,at\,time\,}t\,{\rm is\,
occupied} \\
	0, & {\rm otherwise.}
	\end{array}
	\right.  
	\end{eqnarray}   
Obviously, these variables satisfy the relationships
$R_{i}(t)R_{i}(t)=R_{i}(t)S_{i}(t)=R_{i}(t)$,  
$I_{i}(t)I_{i}(t)=I_{i}(t)S_{i}(t)=I_{i}(t)$,  
$S_{i}(t)S_{i}(t)=S_{i}(t)$, 
$R_{i}(t)\overline{R_{i}(t)}=I_{i}(t)\overline{I_{i}(t)}=0$, and 
$R_{i}(t)I_{i}(t)=R_{i}(t)\overline{S_{i}(t)}
=I_{i}(t)\overline{S_{i}(t)}=0$.  Here $\overline{R_{i}(t)}$ represents 
the conjugate to $R_{i}(t)$ given by $\overline{R_{i}(t)}=1 - R_{i}(t)$, 
with similar definitions for $\overline{S_{i}(t)}$ and 
$\overline{I_{i}(t)}$.  

In order to represent whether 
an immobile atom is advancing or 
retreating {\it at a certain time step} $t$, we define a 
boolean variable $\theta_{i,t}(f)$ at the
$i$th site at time $t$ such that
\begin{eqnarray}
\theta_{i,t}(f) 
=\left\{
	\begin{array}{ll}
	1, &{\rm with\,probability\,}f \\
	0, & {\rm with\,probability\,}(1-f)
	\end{array}
	\right.  
\end{eqnarray}
Thus the term $\theta_{i,t}(\alpha)I_{i}(t)$ represents 
the probability that the $i$th 
site at time $t$ is occupied by an advancing immobile atom, while the term 
$\overline{\theta_{i,t}(\alpha)}I_{i}(t)$ represents the probability that the
$i$th site at time $t$ is occupied by a retreating atom.  Here $\overline{
\theta_{i,t}(\alpha)}$ denotes the conjugate 
of $\theta_{i,t}(\alpha)$, i.e., 
$\overline{\theta_{i,t}(\alpha)} \equiv 1-\theta_{i,t}(\alpha)$. It follows
that 
$\theta_{i,t}(\alpha)\theta_{i,t}(\alpha)=\theta_{i,t}(\alpha)$, 
and 
$\theta_{i,t}(\alpha)\overline{\theta_{i,t}(\alpha)}=0$.  Similarly, in
order to represent whether an advancing or an immobile     
atom can hop into the empty site sandwiched between the two atoms at
a certain time step, another 
boolean variable $\eta_{i,t}(f)$ defined exactly the
same as $\theta_{i,t}(f)$ is introduced with $f=1/2$.  With this, the
factor $\eta_{i,t}(1/2)R_{i}(t)\overline{S_{i+1}(t)}
[\overline{\theta_{i+2,t}(\alpha)
}I_{i+2}(t)]$ represents, for example, 
the probability that the advancing running atom at the $i$th site  
can hop successfully into the ($i+1$)th site at time $t$.  Note that
the two boolean variables $\theta_{i,t}(f)$ and $\eta_{i,t}(f)$ are
statistically uncorrelated.

We study the time evolution of the variables $R_{i}(t)$ and $I_{i}(t)$, 
i.e., we seek the variables $R_{i}(t+1)$ and $I_{i}(t+1)$ as a function 
of quantities at time $t$.  Focusing on $R_{i}(t+1)$, there are various
ways in which the situation at time $t$ affects $R_{i}(t+1)$.  
From the rules A1 and A2, 
a running or an 
advancing immobile atom occupying 
the ($i-1$)th site at time $t$ will hop into the 
empty $i$th site if the ($i+1$)th site is not occupied 
by a retreating immobile atom.  
At the next time step, the ($i-1$)th site will 
become empty while the $i$th site will be occupied by a running 
atom.  This leads to a contribution to 
$R_{i}(t+1)$ of the form
\begin{eqnarray}
[R_{i-1}(t)+\theta_{i-1,t}(\alpha)I_{i-1}(t)] [\overline{S_{i}(t)}]
[\theta_{i+1,t}(\alpha)I_{i+1}(t)   
+ \overline{S_{i+1}(t)}+R_{i+1}(t)].  \nonumber 
\end{eqnarray}  
The three brackets express the conditions on the 
($i-1$)th, $i$th, and ($i+1$)th sites at time $t$, respectively. 
The three terms in the last bracket  
is equivalent to saying that the ($i+1$)th site is not occupied by 
a retreating immobile atom at time $t$.  

The second contribution to $R_{i}(t+1)$ comes from the situation in which 
the ($i+1$)th site {\it is} occupied 
by a retreating immobile atom.  
In this case, the rule A3 leads to another probabilistic event.  
This situation contributes a term to $R_{i}(t+1)$ of the form
\begin{eqnarray}
\eta_{i-1,t}(\frac{1}{2})[R_{i-1}(t)&+&\theta_{i-1,t}(\alpha)I_{i-1}(t)]
[\overline{S_{i}(t)}][\overline{\theta_{i+1,t}(\alpha)}I_{i+1}(t)], 
\nonumber
\end{eqnarray}
which denotes the probability that the atom at the ($i-1$)th site succeeded
in moving forward into the empty $i$th site and became a running atom.
The first factor $\eta_{i-1,t}(\frac{1}{2})$ follows from the rule A3 
as the atoms at the ($i-1$)th and ($i+1$)th sites are equally probable  
to hop into the $i$th site.   

Another contribution comes in 
when the $i$th site is occupied by a running atom, the ($i+1$)th
site is empty and the ($i+2$)th site is occupied by a retreating immobile
atom {\it and} that the retreating atom succeeded 
in hoping back onto the ($i+1$)th site in the process.  In this situation, 
the running atom will stay on the $i$th site at the next time 
step.  This contributes a term 
\begin{eqnarray}
\overline{\eta_{i,t}(\frac{1}{2})} [R_{i}(t)]
[\overline{S_{i+1}(t)}][\overline{\theta_{i+2,t}(\alpha)}I_{i+2}(t)]
\nonumber
\end{eqnarray}           
to $R_{i}(t+1)$, with the brackets expressing the conditions at the 
$i$th, ($i+1$)th, and ($i+1$)th sites.  
A fourth contribution comes from the situation that 
a running atom at the $i$th site is blocked by another running atom 
at the ($i+1$)th site and it gives a term 
\begin{eqnarray}
R_{i}(t) R_{i+1}(t) \nonumber
\end{eqnarray}
to $R_{i}(t+1)$ according to the rule A2. 

Collecting all the four contributions to $R_{i}(t+1)$, we have 
\begin{eqnarray}
R_{i}(t+1)&=&[R+\theta(\alpha)I]_{i-1,t}[\overline{S}]_{i,t}
[\theta(\alpha)I+\overline{S}+R]_{i+1,t} \nonumber \\
&&+[\eta(\frac{1}{2})(R+\theta(\alpha)I)]_{i-1,t}[\overline{S}]_{i,t}
[\overline{\theta(\alpha)}I]_{i+1,t} \nonumber \\
&&+[\overline{\eta(\frac{1}{2})}R]_{i,t}[\overline{S}]_{i+1,t}
[\overline{\theta(\alpha)}I]_{i+2,t}+[R]_{i,t}[R]_{i+1,t}, 
\label{eq:eor}
\end{eqnarray} 
which is a time evolution equation for $R_{i}(t+1)$ in that all the 
quantities on the right hand side are evaluated at time 
$t$.  Note that we have simplified the notations 
so that all the quantities inside a squared bracket are to be evaluated 
at the position and time indicated as subscripts outside the 
bracket.

Similar argument can be carried out for $I_{i}(t+1)$, although the 
analysis is slightly more complicated than the case of $R_{i}(t+1)$.  
The rule A1 states that if the ($i-1$)th site is not 
occupied by a running or an advancing immobile atom, then the retreating
immobile atom at the ($i+1$)th site can hop into the empty $i$th site
deterministically.  This contributes to $I_{i}(t+1)$ a term 
of the form 
\begin{eqnarray}
[\overline{\theta_{i-1}(t)}I_{i-1}(t)+\overline{S_{i-1}(t)}]
[\overline{S_{i}(t)}] [\overline{\theta_{i+1,t}(\alpha)}I_{i+1}(t)],\nonumber
\end{eqnarray}    
where the terms in the first squared bracket 
is equivalent to saying that the ($i-1$)th site
is not occupied by a running or an advancing immobile atom.  

The rule A3 comes into the consideration through various situations. 
If the ($i-1$)th site {\it is} occupied by a running or an advancing
immobile atom, the retreating immobile atom at the ($i+1$)th site
still has half a chance to hop into the empty $i$th site, and contributes 
a term  
\begin{eqnarray}
\overline{\eta_{i-1,t}(\frac{1}{2})} [R_{i-1}(t)&+&\theta_{i-1,t}
(\alpha)I_{i-1}(t)] 
[\overline{S_{i}(t)}] [\overline{\theta_{i+1,t}(\alpha)}I_{i+1}(t)]\nonumber
\end{eqnarray}     
to $I_{i}(t+1)$.  
Note that when an {\it advancing} immobile atom 
{\it fails} to hop forward, it will stay at the site 
and remain immobile.  In this case, an advancing immobile atom  
located at the $i$th site at time $t$ will still be there at time $t+1$ 
contributing a term to $I_{i}(t+1)$ of the form   
\begin{eqnarray}
\left[\overline{\eta_{i,t}(\frac{1}{2})}\theta_{i,t}(\alpha)I_{i}(t)\right]
[\overline{S_{i+1}(t)}] [\overline{\theta_{i+2,t}(\alpha)}I_{i+2}(t)].\nonumber
\end{eqnarray}
An analogous situation arises when the running or advancing
immobile atom located at the ($i-2$)th site succeeded in hopping into the 
empty ($i-1$)th site, leaving a retreating immobile
atom at the $i$th site.  This contributes a term to $I_{i}(t+1)$ as 
\begin{eqnarray}
\eta_{i-2,t}(\frac{1}{2}) [R_{i-2}(t)&+&\theta_{i-2,t}(\alpha)I_{i-2}(t)] 
[\overline{S_{i-1}(t)}] [\overline{\theta_{i,t}(\alpha)}I_{i}(t)].\nonumber
\end{eqnarray}

Blocking by atoms in the nearest neighboring sites contributes the 
following terms.  
If an advancing immobile atom at the $i$th site is blocked by the 
an atom at the ($i+1$)th site, or a retreating immobile atom 
is blocked by an atom  at the ($i-1$)th site, the state of the $i$th
site at time $t+1$ remains to be immobile.  These two terms in 
$I_{i}(t+1)$ are represented by
$[\theta_{i,t}(\alpha)I_{i}(t)] S_{i+1}(t)+ S_{i-1}(t)
[\overline{\theta_{i,t}(\alpha)}I_{i}(t)]$. 
Finally according to rule A2, a term $R_{i}(t) I_{i+1}(t)$ in $I_{i}(t+1)$
arises from the blocking of a running atom by an immobile atom.

Collecting all the contributions to $I_{i}(t+1)$, we have 
\begin{eqnarray}
I_{i}(t+1)&=&[\overline{\theta(\alpha)}I+\overline{S}+\overline{
\eta(\frac{1}{2})}(R+\theta(\alpha)I)]_{i-1,t}[\overline{S}]_{i,t}[
\overline{\theta(\alpha)}I]_{i+1,t} \nonumber \\
&&+[\overline{\eta(\frac{1}{2})}\theta(\alpha)I]_{i,t}[\overline{S}]_{i+1,t}
[\overline{\theta(\alpha)}I]_{i+2,t} \nonumber\\
&&+[\eta(\frac{1}{2})(R+\theta(\alpha)I)]_{i-2,t}[\overline{S}]_{i-1,t}
[\overline{\theta(\alpha)}I]_{i,t} \nonumber \\
&&+[\theta(\alpha)I]_{i,t}[S]_{i+1,t}+[S]_{i-1,t}[\overline
{\theta(\alpha)}I]_{i,t} \nonumber \\
&&+[R]_{i,t}[I]_{i+1,t}.  \label{eq:eoi}
\end{eqnarray}
Equations (\ref{eq:eor}) and (\ref{eq:eoi}) can be used to compute the 
time 
evolution of the mobility of the system and the spatial 
averages of the products of different combinations of the 
state variables defined on the same or neighboring sites.

The mobility $B(t)$ at time $t$, i.e.,  the fraction of atoms in 
the running state at time $t$, can be expressed in terms of $R_{i}(t)$ as  
\begin{eqnarray}
B(t)\equiv\frac{1}{N}\sum_{i}R_{i}(t)= \frac{1}{\rho} \langle R_{i}(t) 
\rangle, \label{eq:fdom}
\end{eqnarray} 
where $\langle \cdots \rangle \equiv 1/N \sum_{i} (\cdots)$ is the  
spatial average of the quantity concerned over the system.  It follows that 
$B(t+1)=\frac{1}{\rho}\langle R_{i}(t+1)\rangle$. 
Making use of the expression for $R_{i}(t+1)$ in terms of 
quantities at time $t$ 
given by Eq.(\ref{eq:eor}), the mobility at time $t+1$ can be expressed  
in terms 
of spatial averages involving strings of upto 
three neighboring sites at time $t$. 
This gives 
\begin{eqnarray}
B(t+1)&=&[\alpha \langle R0I \rangle_{t}+\alpha^{2}\langle I0I\rangle_{t}
+\langle R00\rangle_{t}+\alpha\langle I00\rangle_{t}+\langle R0R\rangle_{t}
+\alpha\langle I0R\rangle_{t}] \nonumber\\
&&+\frac{1-\alpha}{2}[\langle R0I\rangle_{t}+\alpha\langle I0I\rangle_{t}]
+\frac{1-\alpha}{2}\langle R0I\rangle_{t} +\langle RR\rangle_{t} .\label{eq:b1}
\end{eqnarray}
For simplicity, we write the
spatial averages at time $t$ as $\langle \cdots \rangle_{t}$ and  
express the strings of neighboring sites in order from left to right.  
We use the symbol  
"0" to denote an empty site or $\overline{S}$.  
For example, $\langle R0I \rangle_{t}$ implies counting the strings 
of neighboring sites with a running atom on the left and an immobile 
atom on the right with an empty site in between over the system at
time $t$.  The prefactors resulted from the fact that 
the spatial averages of the boolean 
variable $\theta(\alpha)$ (or its conjugate)  
survives with a probability 
$\alpha$ (or $1-\alpha$).  Similarly, $\eta(1/2)$ survives  
with probability $1/2$ under averaging.  
Thus, the term $\alpha \langle R0I\rangle_{t}$ comes from 
the average $\langle R0\theta
(\alpha)I\rangle_{t}$.  
Noting that  
$S_{i}(t)+\overline{S_{i}(t)}=R_{i}(t)+I_{i}(t)+\overline{S_{i}(t)}=1$, 
we have 
\begin{eqnarray}
B(t+1)&=&\frac{1}{\rho}\left[ \langle RR\rangle_{t}+\langle R0\rangle_{t}
+\alpha\langle I0\rangle_{t}-\alpha\frac{1-\alpha}{2}\langle I0I\rangle_{t}
\right]. \label{eq:mt1}
\end{eqnarray}
Equation (\ref{eq:mt1}) is an exact expression for $B(t+1)$ in terms of 
quantities evaluated at time $t$.  
In order to proceed, we 
write down the evolution equation for the spatial averages 
on the right hand side of Eq.(\ref{eq:mt1}). Obviously iterating the 
equations  
backward in time gives terms involving longer strings of neighboring
sites and hence longer spatial correlations.
To close the set of equations, a decoupling scheme retaining spatial 
averages involving two neighboring sites  is invoked.  The set 
of equations can then be treated as a dynamical system.  The 
fixed points of the equations then give the results corresponding to 
the long time limit.

To treat the system analytically, we decouple the term 
$\langle I0I \rangle_{t}$ in Eq.(\ref{eq:mt1}) into products of averages 
involving two neighboring sites, i.e., $\langle I0I \rangle_{t} \approx
\langle I0 \rangle_{t} \langle 0I \rangle_{t} /(1-\rho)$, where 
$1-\rho = \langle 0 \rangle_{t}$ is the probability of finding an 
empty site\cite{kwong1}.  With this approximation, Eq.(\ref{eq:mt1}) 
becomes 
\begin{eqnarray}
B(t+1)&=& \frac{1}{\rho}\left[ \langle  RR\rangle_{t}+\langle R0\rangle_{t}
+\alpha \langle I0\rangle_{t}
-\alpha \frac{1-\alpha}{2}\frac{\langle I0\rangle_{t}\langle 0I\rangle_{t}}
{1-\rho}\right]. \label{eq:mt1g} 
\end{eqnarray}
With the four variables $R$, $I$, $S$ and $\overline{S}$, a total of 
sixteen two-site spatial averages can be formed, among which four of them 
can be chosen to be independent.  
We choose the independent spatial averages to be 
$\langle  RR\rangle_{t} ,\langle RI
\rangle_{t}, \langle IR\rangle_{t}$ and $\langle II\rangle_{t}$.  The other 
two-site averages are related through 
\begin{eqnarray}
\langle R1\rangle_{t}&=&\langle RR\rangle_{t}+\langle RI\rangle_{t}, \nonumber \\
\langle 1R\rangle_{t}&=&\langle RR\rangle_{t}+\langle IR\rangle_{t} ,\nonumber \\
\langle I1\rangle_{t}&=&\langle IR\rangle_{t}+\langle II\rangle_{t},\nonumber \\
\langle 1I\rangle_{t}&=&\langle RI\rangle_{t}+\langle II\rangle_{t},\nonumber \\
\langle R0\rangle_{t}&=&\rho B(t) -\langle R1\rangle_{t}, \nonumber \\
\langle I0\rangle_{t}&=&\rho(1-B(t))-\langle I1\rangle_{t}, \nonumber \\
\langle 0R\rangle_{t}&=&\rho B(t) -\langle 1R\rangle_{t} ,\nonumber \\
\langle 0I\rangle_{t}&=&\rho (1-B(t)) -\langle 1I\rangle_{t}, \nonumber \\
\langle 01\rangle_{t}&=&\rho -\langle R1\rangle_{t}-\langle I1\rangle_{t} ,
\nonumber \\
\langle 10\rangle_{t}&=&\langle 01\rangle_{t} ,\nonumber \\
\langle 00\rangle_{t}&=&1-\rho-\langle 01\rangle_{t},     \nonumber\\
\langle 11\rangle_{t}&=&\langle I1\rangle_{t} + \langle R1\rangle_{t},
\label{eq:dsp} 
\end{eqnarray}
where ``$1$" represents an occupied site regardless the character of 
the atom.  Using $\langle RR\rangle_{t+1} = \langle R_{i}(t+1) R_{i+1}(t+1) 
\rangle$ and Eq.(\ref{eq:eor}) for $R_{i}(t+1)$ and $R_{i+1}(t+1)$, we have 
\begin{eqnarray}
\langle RR\rangle_{t+1} &=&\langle RRR\rangle_{t}+\frac{1-\alpha}{2}\langle
RR0I\rangle_{t}+\langle R0RR\rangle_{t} +\alpha\langle I0RR\rangle_{t} \nonumber\\
&&+\frac{1-\alpha}{2}[\langle R0R0I \rangle_{t}+\alpha\langle I0R0I\rangle_{t}.
\label{eq:rrg}
\end{eqnarray}
To make the approximation self-consistent, 
we invoke the decoupling scheme and retaining spatial averages involving 
no more than two sites, we have 
\begin{eqnarray}
\langle RR \rangle_{t+1}&=&\frac{\langle RR\rangle_{t}^{2}}{\rho B(t)}
+\frac{\langle RR\rangle_{t}}{(1-\rho)\rho B(t)}\left\{\frac{1-\alpha}{2}
\langle R0\rangle \langle 0I\rangle_{t}+\langle 0R\rangle_{t}[\langle
R0\rangle_{t} +\alpha\langle I0\rangle_{t}]\right\} \nonumber \\
&&+\frac{1-\alpha}{2}\frac{\langle R0\rangle_{t}\langle 0R\rangle_{t}\langle
0I\rangle_{t}}{(1-\rho)^{2}\rho B(t)}[\langle R0\rangle_{t}+\alpha
\langle I0\rangle_{t}].  \label{eq:rrgd} 
\end{eqnarray} 
Similarly, for $\langle RI \rangle_{t+1}$, $\langle IR \rangle_{t+1}$, and 
$\langle II \rangle_{t+1}$, we obtain after decoupling
\begin{eqnarray}
\langle RI\rangle_{t+1}&=&\frac{\langle RR\rangle_{t}\langle RI\rangle_{t}}
{\rho B(t)}+\frac{1-\alpha}{1-\rho}\langle 0I\rangle_{t}\left[ \langle
R0\rangle_{t}+\frac{\alpha}{2}\langle I0\rangle_{t}\right]
+\frac{1}{1-\rho}[\langle R0\rangle_{t}+\alpha \langle I0\rangle_{t}] \nonumber \\
&&\cdot \left\{
\frac{\langle 0R\rangle_{r}\langle RI\rangle_{t}}{\rho B(t)} \right.
+\frac{1-\alpha}{1-\rho}\langle 00\rangle_{t}\langle 0I\rangle_{t}
+\alpha \frac{\langle 0I\rangle_{t}\langle I1\rangle_{t}}{\rho (1-B(t))}  \nonumber \\
&&\left.+\frac{\alpha}{2} \frac{1-\alpha}{1-\rho}\frac{\langle 0I\rangle_{t}^{2}\langle
I0\rangle_{t}}{\rho (1-B(t))} \right\}, \label{eq:rigd}\\
\langle IR\rangle_{t+1}&=&\frac{1}{\rho B(t)}\left[\langle RR\rangle_{t}
+\frac{1-\alpha}{2(1-\rho)}\langle R0\rangle_{t}\langle I0\rangle_{t} 
\right] \left\{ \alpha \langle IR\rangle_{t}+(1-\alpha)\frac{\langle 1I\rangle_{t}
\langle IR\rangle_{t}}{\rho (1-B(t))}\right. \nonumber \\
&&\left.+\frac{1-\alpha}{2(1-\rho)}\frac{\langle 0I\rangle_{t}\langle IR\rangle_{t}}
{\rho (1-B(t))} \right\}, \\
\langle II\rangle_{t+1}&=&(1-\alpha)[\langle RI\rangle_{t}+\alpha \langle
II\rangle_{t} ]+\frac{\alpha}{2}\frac{1-\alpha}{1-\rho}\langle 0I\rangle_{t}
\langle I0\rangle_{t} +\frac{\alpha}{\rho B(t)}\langle RI\rangle_{t}\langle
IR\rangle_{t} \nonumber \\
&&+(1-\alpha)^{2}\frac{\langle 1I\rangle_{t}\langle II\rangle_{t}}{\rho (1-B(t))}+\alpha \left[
\frac{\langle RI\rangle_{t}\langle I1\rangle_{t}}{\rho (1-B(t))}+
\alpha \frac{\langle II\rangle_{t}\langle I1\rangle_{t}}{\rho (1-B(t))}\right] \nonumber \\
&&+\frac{(1-\alpha)^{2}}{1-\rho}\frac{\langle 0I\rangle_{t} \langle
1I\rangle_{t}\langle I0\rangle_{t}}{\rho (1-B(t))}+\frac{1-\alpha}{\rho B(t)}
\frac{\langle RI\rangle_{t}\langle 1I\rangle_{t}\langle IR\rangle_{t}}{
\rho (1-B(t))} \nonumber \\
&&+\alpha(1-\alpha)\frac{\langle 1I\rangle_{t}\langle II\rangle_{t}
\langle I1\rangle_{t}}{\rho^{2}(1-B(t))^{2}}+\frac{\alpha}{2}\frac{1-\alpha}{1-
\rho}\frac{\langle I0\rangle_{t}\langle 0I\rangle_{t}}{
\rho^{2}(1-B(t))^{2}} \nonumber \\
&&\cdot \left[\langle RI\rangle_{t}
 +\alpha \langle II\rangle_{t}] \right.
+\frac{(1-\alpha)^{2}}{2(1-\rho)}\frac{\langle 0I\rangle_{t}
\langle II\rangle_{t}}{\rho (1-B(t))} 
[\langle R0\rangle_{t}+
\alpha \langle I0\rangle_{t}] \nonumber \\
&&+\frac{1}{2}\left( \frac{1-\alpha}{1-\rho}\right) ^{2}
\frac{\langle 0I\rangle_{t}^{2}\langle I0\rangle_{t}}{\rho (1-B(t))}[\langle 
R0\rangle_{t}+\alpha \langle I0\rangle_{t}] 
+\frac{\alpha}{2}\frac{(1-\alpha)^{2}}{1-\rho} \nonumber \\
&&\cdot \frac{\langle 0I \rangle_{t}
\langle 1I\rangle_{t}\langle II\rangle_{t}\langle I0\rangle_{t}}{\rho^{2}
(1-B(t))^{2}} +\frac{1-\alpha}{2(1-\rho)\rho B(t)}\frac{\langle 0I\rangle_{t}
\langle IR\rangle_{t}\langle RI\rangle_{t}}{\rho (1-B(t))}[\langle R0 \rangle_{t}
\nonumber \\
&&+\alpha \langle I0\rangle_{t}]+\frac{\alpha}{2}\frac{1-\alpha}{1-\rho}
\frac{\langle I1\rangle_{t} \langle
0I\rangle_{t}\langle II\rangle_{t}}{\rho^{2}(1-B(t))^{2}}[\langle R0\rangle_{t}
+\alpha \langle I0\rangle_{t}] \nonumber \\
&&+\frac{\alpha}{4}\left(\frac{1-\alpha}{1-\rho}\right)^{2}\frac{
\langle I0 \rangle_{t}\langle 0I\rangle_{t}^{2}\langle II\rangle_{t}}{
\rho^{2} (1-B(t))^{2}}[\langle R0\rangle_{t}+\alpha \langle I0\rangle_{t}]. 
\label{eq:iigd}
\end{eqnarray}

Equations (\ref{eq:mt1g}) and (\ref{eq:rrgd})- (\ref{eq:iigd})
form a set of five equations for $B$, $\langle RR
\rangle$, $\langle RI \rangle$, $\langle IR \rangle$, and $\langle II \rangle$.
These equations form a five dimensional dynamical system.  To compare with
simulation data, we solve for the stable fixed points numerically.  Results
for the mobility are shown in Fig.1 as solid lines for different values
of $\alpha$ and $\rho$.  The analytical results are in good agreement with
numerical data, showing that the decoupling approximation is sufficient to
capture the essential features of the model.  We have also checked the
results of the two-site and three-site spatial averages numerically and
analytically and it is found that the decoupling scheme gives
qualitatively correct results for the spatial averages.  Our method 
represents a systematic way of deriving mean field theories from 
microscopic consideration by following the time evolution of the system.  
The decoupling scheme of retaining two-site spatial averages is the 
minimal procedure to obtain qualitatively correct mobility and spatial 
averages involving longer strings of sites\cite{kwong1}.   

The $\alpha =1$ case deserves further discussion.  
For $\alpha=1$,  Eq.(\ref{eq:mt1})
gives 
\begin{equation}
B(t+1)=\frac{1}{\rho}[\langle 10\rangle_{t}+\langle RR\rangle_{t}],
\label{eq:mt1a1}
\end{equation}
where $\langle 10\rangle_{t} = \langle R0\rangle_{t} + \langle I0\rangle_{t}$.
The spatial average $\langle RR \rangle_{t+1}$ can be obtained by
setting $\alpha = 1$ in Eq.(\ref{eq:rrgd}) to get
\begin{equation}
\langle RR\rangle_{t+1}=\frac{\langle 10\rangle_{t}
\langle 0R\rangle_{t}\langle RR\rangle_{t}}{(1-\rho)\rho B(t)}
+\frac{\langle RR\rangle_{t}^{2}}{\rho B(t)}. 
\label{eq:rr2s}
\end{equation}
To form a closed set of equations, we work out the spatial averages
$\langle 10\rangle_{t+1}$ and $\langle 0R\rangle_{t+1}$ within the
approximation of retaining two-site correlations to get
\begin{equation}
\langle 10\rangle_{t} = \rho - (1-\frac{\langle 10\rangle_{t}}{\rho})
(\rho-\langle 10\rangle_{t} + \frac{\langle 10 \rangle_{t}^{2}}{1-\rho}),
\end{equation}
and
\begin{equation}
\langle 0R\rangle_{t+1}=\langle 10\rangle_{t}
+\frac{\langle 0R\rangle_{t}\langle RR\rangle_{t}}{\rho B(t)}
-\frac{\langle 10\rangle_{t}\langle 0R\rangle_{t}\langle RR\rangle_{t}}{
(1-\rho)\rho B(t)}, 
\label{eq:0r2s}
\end{equation}
where we have used the relations stated in Eq.(\ref{eq:dsp}).  The fixed 
points satisfy
$B(t+1) = B(t) \equiv B$, $\langle 10\rangle_{t+1} = \langle 10\rangle_{t}
\equiv y$, $\langle RR\rangle_{t+1} = \langle RR\rangle_{t} \equiv z$,
and $\langle 0R\rangle_{t+1} = \langle 0R\rangle_{t} \equiv w$.  Hence, 
they  satisfy the simltaneous equations
\begin{eqnarray}
\rho B& =& y+z, \\
y& =& \rho - (1-\frac{y}{\rho}) (\rho-y + \frac{y^{2}}{1-\rho}), \label{eq:y}\\
z& =& \frac{z^{2}}{\rho B} + \frac{wyz}{(1-\rho)\rho B},\label{eq:z} \\
w&=& y + \frac{wz}{\rho B} - \frac{wyz}{(1-\rho)\rho B}.\label{eq:w}
\end{eqnarray}
From Eq.(\ref{eq:y}), the stable fixed point for $y$ is given by $y=\rho$ for
$\rho < 1/2$ and $y=1-\rho$ for $\rho \geq 1/2$.   From Eq.(\ref{eq:z}),
$z=0$ is a stable fixed point for $\rho <1/2$ and $z=\rho B -w$ is a
stable fixed point for $\rho \geq 1/2$, where we have used the results
for $y$.  It is important to note that Eq.(\ref{eq:w}) for $w$ becomes 
redundant.
Hence the situation is that we have three equations with four unknowns.
The values of $B$ and $z$ are governed by the linear relation
\begin{equation}
z = \rho B - (1-\rho). \label{eq:lr}
\end{equation}
Any values of $x\in [(1-\rho)/\rho,1]$ and $z\in [0,2\rho-1]$
satisfying Eq.(\ref{eq:lr}) is a solution to the system of equations.  
Thus for $\alpha =1$, the
mobility $B = 1$ for $\rho < 1/2$ and $B$ lies in the range
$[(1-\rho)/\rho,1]$ for $\rho \geq 1/2$ with the precise value
depending on the initial condition, in agreement with    
numerical results.  The value $(1-\rho)/\rho$
shown in Fig.1 corresponds
to the initial condition of all the atoms being immobile.

\subsection{Model B}
The new parameter $\gamma$ introduced in rule B1 is the probability 
that an atom in the running state changes into the immobile state in 
a time step.  To carry out analytical treatments similar to those in 
Model A, it is convenient to divide each time interval into two 
halves.  In the first half of a time step, rule B1 applies and 
the parameter $\gamma$ is effective; while in the second half of the 
time step, rules A1 and A2 apply.  Introducing a Boolean variable 
$\zeta_{i,t}(\gamma)$ analogous to, but statistically independent of, 
$\theta_{i,t}(\gamma)$ and $\eta_{i,t}$, the 
variables $R_{i}(t)$ and $I_{i}(t)$ evolve in the first half of the 
time step as:
\begin{equation}
R_{i}(t+\frac{1}{2})=\overline{\zeta_{i,t}(\gamma)}R_{i}(t) ,\label{eq:1hr}
\end{equation}
and
\begin{equation}
I_{i}(t+\frac{1}{2})=I_{i}(t)+\zeta_{i,t}(\gamma)R_{i}(t). \label{eq:1hi}
\end{equation}
The time evolution in the second half of the time step is given by 
Eqs.~(\ref{eq:eor}) and (\ref{eq:eoi}), with the quantities on the 
right hand side of the equations corresponding to those evaluated 
at $t+1/2$.  Combining the evolution in the two halves of a time step, 
we finally arrive at 
\begin{eqnarray}
R_{i}(t+1)&=&\{[\overline{\zeta(\gamma)}+\theta(\alpha)\zeta(\gamma)]R+
\theta(\alpha)I\}_{i-1,t}[\overline{S}]_{i,t}[\overline{S}
+(\overline{\zeta(\gamma)} \nonumber \\
&&+\theta(\alpha)\zeta(\gamma))R 
+\theta(\alpha)I]_{i+1,t}+\{\eta(\frac{1}{2})[\overline{\zeta(\gamma)}+
\theta(\alpha)\zeta(\gamma)]R \nonumber \\
&&+\eta(\frac{1}{2})\theta(\alpha)I\}_{i-1,t}[\overline{S}]_{i,t}[\overline{
\theta(\alpha)}\zeta(\gamma)R+\overline{\theta(\alpha)}I]_{i+1,t} \nonumber\\
&&+[\overline{\eta(\frac{1}{2})}\overline{\zeta(\gamma)}R]_{i,t}[\overline{S}
]_{i+1,t}[\overline{\theta(\alpha)}\zeta(\gamma)R+\overline{\theta(\alpha)}I
]_{i+2,t} \nonumber \\
&&+[\overline{\zeta(\gamma)}R]_{i,t}[\overline{\zeta(\gamma)}R]_{i+1,t}, 
\label{eq:eorb} 
\end{eqnarray}
and
\begin{eqnarray}
I_{i}(t+1)&=&\{\overline{S}+[\zeta(\gamma)\overline{\theta(\alpha)}+
\overline{\eta(\frac{1}{2})}\overline{\zeta(\gamma)}+\overline{
\eta(\frac{1}{2})}\theta(\alpha)\zeta(\gamma)]R \nonumber \\
&&+[\overline{\theta(\alpha)}+\overline{\eta(\frac{1}{2})}\theta(\alpha)]I
\}_{i-1,t}[\overline{S}]_{i,t}[\overline{\theta(\alpha)}\zeta(\gamma)R \nonumber \\
&& +\overline{\theta(\alpha)}I]_{i+1,t}+[\overline{\eta(\frac{1}{2})}
\theta(\alpha)\zeta(\gamma)R+\overline{\eta(\frac{1}{2})}\theta(\alpha)I]_{i,t} 
\nonumber \\
&&[\overline{S}]_{i+1,t}[\overline{\theta(\alpha)}\zeta(\gamma)R+\overline{
\theta(\alpha)}I]_{i+2,t} +\{ [\eta(\frac{1}{2})\overline{\zeta(\gamma)} \nonumber \\
&&+\eta(\frac{1}{2})\theta(\alpha)\zeta(\gamma)]R+\eta(\frac{1}{2})\theta(\alpha)
I\}_{i-2,t}[\overline{S}]_{i-1,t} \nonumber \\
&&\cdot[\overline{\theta(\alpha)}\zeta(\gamma)R+\overline{\theta(\alpha)}I
]_{i,t} +[\theta(\alpha)\zeta(\gamma)R \nonumber \\
&&+\theta(\alpha)I]_{i,t}[S]_{i+1,t} +[S]_{i-1,t}[\overline{\theta(\alpha)}
\zeta(\gamma)R \nonumber \\
&&+\overline{\theta(\alpha)}I]_{i,t} +[\overline{\zeta(\gamma)}R]_{i,t}
[\zeta(\gamma)R +I]_{i+1,t}.  \label{eq:eoib}
\end{eqnarray}
Equations (\ref{eq:eorb}) and (\ref{eq:eoib}) are the time evolution 
equations relating $R_{i}(t+1)$ and $I_{i}(t+1)$ to quantities at time 
$t$.  They play exactly the same role as Eqs.~(\ref{eq:eor}) 
and ~(\ref{eq:eoi})
in model A. 

It is then straightforward to carry out the same treatment for model B as 
in model A, and we simply outline the key steps in the following 
discussion.  Following the same steps leading to 
Eq.~(\ref{eq:mt1}), the mobility $B(t+1)$ at time $t+1$ for model B is given 
by 
\begin{eqnarray}
B(t+1)&=&\frac{1}{\rho}\{ (1-\gamma)^{2}\langle RR\rangle_{t} +
(1-\gamma+\alpha \gamma)\langle R0\rangle_{t}+\alpha \langle I0\rangle_{t} \nonumber \\
&&-\frac{\alpha}{2} (1-\alpha )[\langle I0I \rangle_{t} +\gamma(\langle R0I
\rangle_{t} +\langle I0R\rangle_{t}) \nonumber \\
&&+\gamma^{2}\langle R0R\rangle_{t}]\}. \label{eq:mt1b}
\end{eqnarray}
Equation (\ref{eq:mt1b}) is the generalization of Eq.(\ref{eq:mt1}) to 
model 
B.  It reduces to Eq.(\ref{eq:mt1}) for $\gamma =0$.  Employing a 
decoupling approximation to retain only spatial averages involving 
upto two nearest neighboring sites, Eq.(\ref{eq:mt1b}) becomes
\begin{eqnarray}
B(t+1) &=& \frac{1}{\rho}\{ (1-\gamma)^{2}\langle RR\rangle_{t}
+(1-\gamma +\alpha \gamma )\langle R0 \rangle_{t} +\alpha \langle I0 
\rangle_{t} \nonumber \\
&&-\frac{\alpha}{2}\frac{1-\alpha}{1-\rho}[\langle I0\rangle_{t}
+\gamma \langle R0\rangle_{t} ][\langle 0I\rangle_{t} +\gamma \langle 
0R\rangle_{t} ] \}.    
\end{eqnarray}
To close the set of equations, we construct the time evolution equations 
for the spatial averages $\langle RR\rangle$, $\langle RI\rangle$, 
$\langle IR\rangle$, and $\langle II\rangle$.  The other spatial 
averages can be constructed from these four averages.  In the presence 
of the parameter $\gamma$, the resultant equations are more complicated 
than Eqs.(13)-(16) in model A.  This set of equations forms a dynamical 
system.  The stable fixed point corresponding to the mobility and spatial 
averages in the long time limit can be readily solved numerically.  Results
for the mobility $B$ in the steady state are shown as solid lines in
Figs. 2 and 3 for two different values of atomic concentration $\rho$.  The 
theoretical results obtained within the decoupling approximation capture 
all the essential features of the numerical data.  It is observed that the 
theoretical results are consistently slightly greater than the numerical 
data.  The discrepancies come from the decoupling scheme.  If the 
decoupling approximation is extended to retain spatial averages involving 
upto three neighboring sites, for which the calculations are much 
more involved, the results are in better agreement with numerical 
data\cite{kwong2}.  It is, however, important to stress that the essential 
physics is captured within the two-site decoupling approximation.

The particular case of $\alpha =1$ can be treated in a way analogous to
that in model A.  It is found that  
for $\gamma\in (0,1]$, the stable attractors give the mobility
\begin{eqnarray}
B(t\rightarrow \infty)= \left\{
\begin{array}{lll}
1&{\rm if}&\rho<1/2 \\
\frac{1-\rho}{\rho}&{\rm if}&\rho\geq 1/2 
\end{array}
\right.  ,
\end{eqnarray}
together with the spatial averages
\begin{eqnarray}
\langle 10\rangle_{t\rightarrow \infty} &=& 
\langle 0R\rangle_{t\rightarrow \infty} \nonumber \\
&=& \left\{
\begin{array}{lll}
\rho&{\rm if}&\rho<1/2 \\
1-\rho&{\rm if}&\rho\geq 1/2 
\end{array}
\right. , 
\end{eqnarray}
and
\begin{equation}
\langle RR\rangle_{t\rightarrow \infty} =0.
\end{equation}
For $\alpha =1$ and $\gamma \neq 0$, the time evolution of the state at a
site depends on the states of the nearest neighboring sites only and hence 
the decoupling scheme retaining only two-site spatial averages is good.  
Results so obtained are in exact agreement with numerical data. 
It should be noted that for
$\gamma=0$ and $\alpha =1$, $B=1$ for $\rho <1/2$ and $B$ lies in the range
$[(1-\rho)/\rho,1]$ for $\rho \geq 1/2$ with the precise value depending
on the initial condition as discussed in the previous subsection. 

\section{Summary}
In summary,  we have proposed three CA models defined in terms of
parallel updating rules analogous to 
 the three models recently studied by Braun {\it et 
al.} \cite{braun3} which are defined in terms of 
sequentially updating rules.  These models are of increasing complexity so 
as to model the generalized FK models proposed recently within the context
of tribology.  
The first 
model (Model A) involves atoms in two different dynamical states, i.e., 
running
and immobile, subjected to an external field parametrized by 
$\alpha$.  Atoms in the running state tend to hop along the field
direction
while atoms in the immobile state may bounce backward.  The second 
model (Model B) involves spontaneous transition of
 atoms from the running state to the immobile state in addition to 
 the rules of Model 
 A.  The third model
 (Model C) takes into account of the crowding effect of the system as 
 well.  Results of numerical simulations indicate that 
 the mobility, which is 
 defined as the fraction of atoms in the running state, as a 
function of $\alpha$ for the three models 
 exhibit phase transitions, jammings and 
 hysteresis.  The proposed CA 
models have the advantage that they can be treated as dynamical systems  
and analysed by the microscopic approach previously applied to 
analogous models of traffic flow problems \cite{wang}. 
For Models A and B in which the state of a site is affected by the 
nearest and next
nearest neighbours, time evolution equations of the state of the sites 
can be explicitly written down.  The mobility of the system at time $t+1$ 
can be expressed in terms of spatial averages at time $t$ involving not more 
than
three sites.  By invoking an approximation which decouples the 
spatial averages into products of spatial averages involving upto 
two neighboring sites as well as deriving the evolution 
equations of
the two-site spatial averages, a closed set of equations forming  
a nonlinear dynamical mapping can be established.  The behavior in the long 
time limit can be found by studying the stable fixed point of the 
mapping.  The fixed point
of the mapping can be found numerically.  In the
special case that $\alpha=1$ and $\gamma \neq 0$, the decoupling scheme 
is exact and analytical solutions can be found. For the whole range 
of possible values of the parameters in the models, 
the present approach yields satisfactory results when compared with 
simulation data.  Our analytical approach represents a 
systematic way to derive approximations of increasing accuracy starting 
from a microscopic point of view by following the time evolution of the 
system.  It is expected that better agreement with numerical results  
can be obtained by retaining spatial averages involving a long string 
of neighboring sites.   

\newpage
\begin{center}
{\bf ACKNOWLEDGMENTS}
\end{center} 

B.H.W. acknowledges the support from the Chinese National Basic Research 
Climbing-up Project ``Nonlinear Science" and the National Natural Foundation 
in China.  P.M.H acknowledges the support of the Research Grants Council 
(RGC) of the Hong Kong SAR Government under Grant No. CUHK 4191/97P.  The 
work of B.H. and B.H.W. was also supported in part by grants from the 
Hong Kong Research Grants Council (RGC) and the Hong Kong Baptist 
University Faculty Research Grants (FRG).  We thank Prof. O.M. Braun for 
useful discussions.

\newpage
\begin{figure}
\caption{The mobility $B$, which is defined as the fraction of atoms 
in the running state, of Model A in the long time limit as a function of 
the external biasing parameter $\alpha$ for different values of the 
concentration of atoms $\rho$.  The symbols are numerical data and the 
solid lines are results obtained by invoking the decoupling scheme 
in the analytical approach discussed in the text.}  
\end{figure}
\begin{figure}
\caption{The mobility $B$ of Model B in the long time limit as a function 
of the parameter $\alpha$ for different values of the parameter $\gamma$ 
at fixed concentration $\rho =0.4$.  The symbols are numerical results 
and the solid lines are results obtained by invoking the decoupling 
scheme.  Results are typical of those for $\rho < 1/2$.}
\end{figure}
\begin{figure}
\caption{The mobility $B$ of Model B in the long time limit as a function 
of the parameter $\alpha$ for different values of the parameter $\gamma$ 
at fixed concentration $\rho =0.6$.  The symbols are numerical results 
and the solid lines are results obtained by invoking the decoupling 
scheme.  Results are typical of those for $\rho > 1/2$.}
\end{figure}
\begin{figure}
\caption{The mobility $B$ of Model C in the long time limit as a function 
of the parameter $\alpha$ for different values of the concentration $\rho$. 
The parameter $\gamma_{0}$ is taken to be $10^{-5}$.  Results are 
typical for those with $\gamma_{0} \approx 0$.  Hysteresis in the 
mobility is observed for $\rho \leq 1/2$.}
\end{figure}

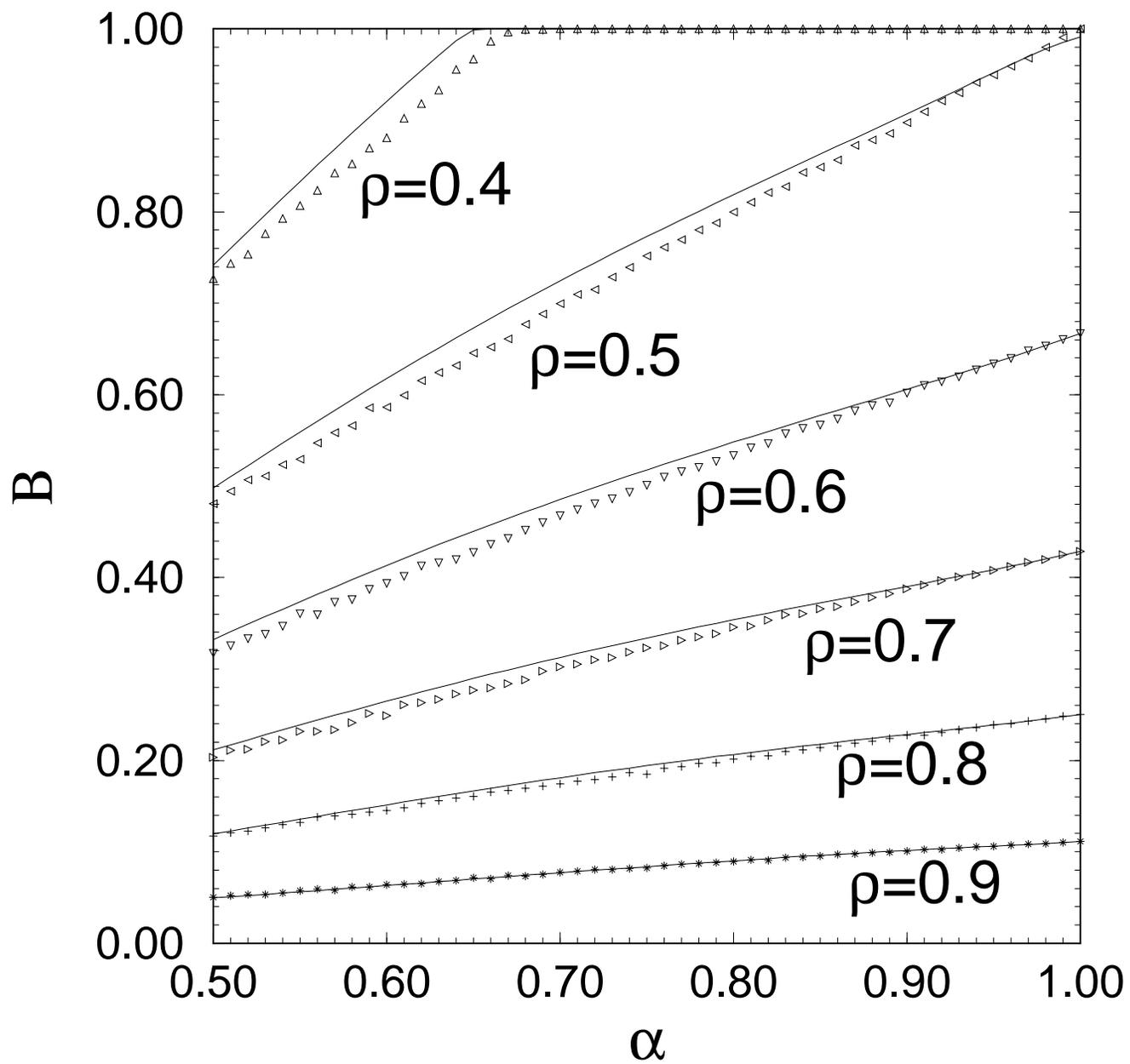
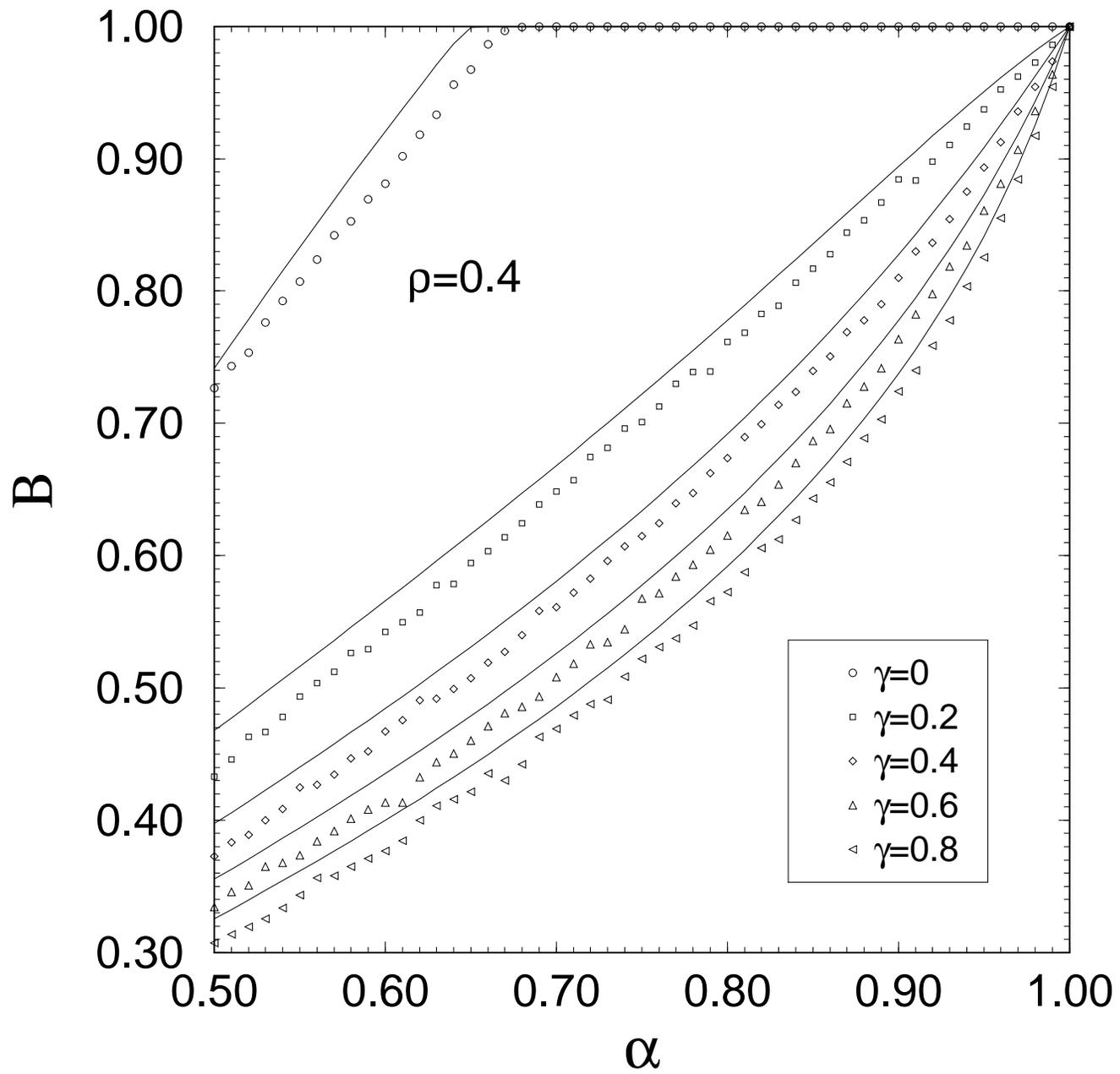
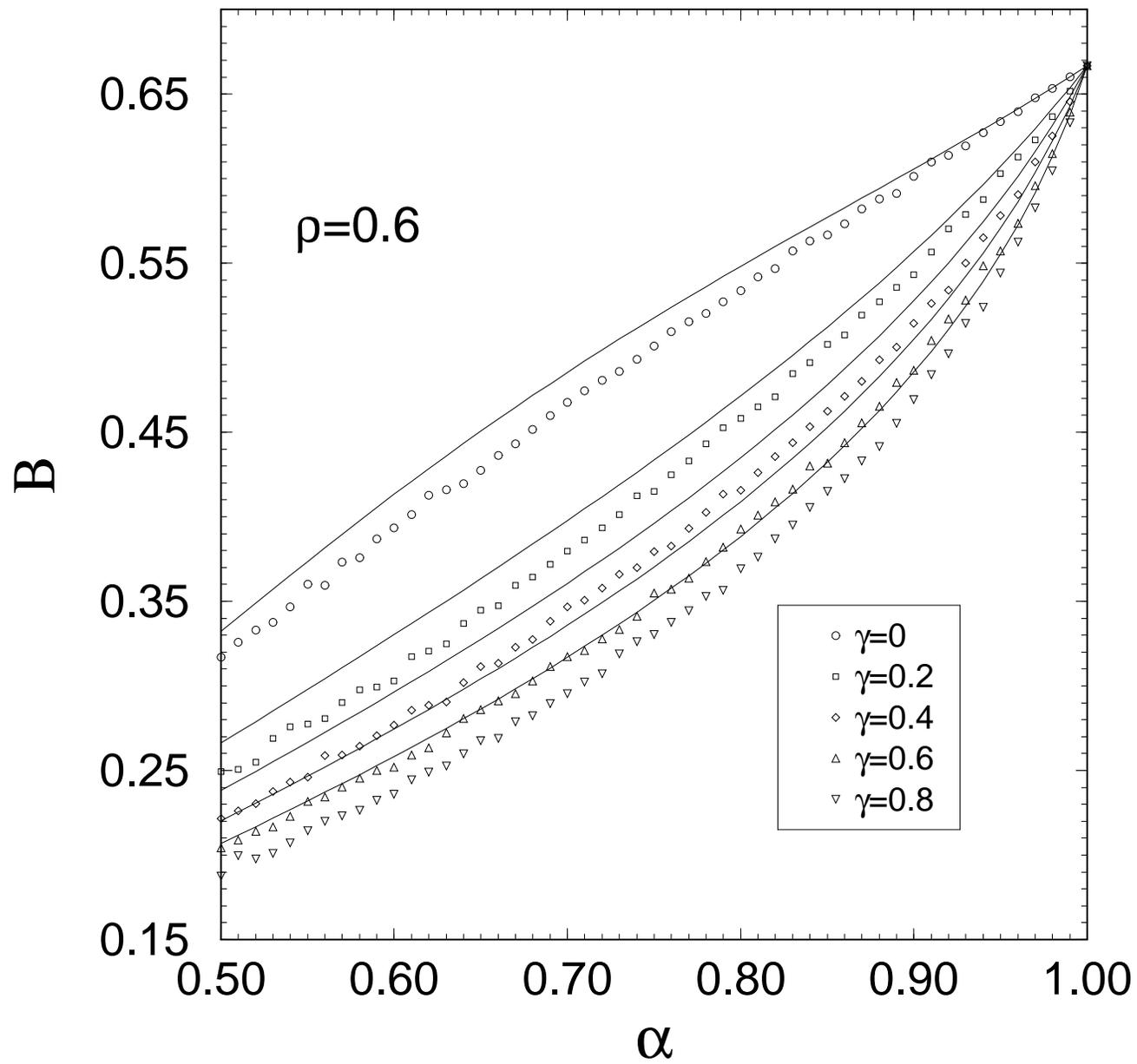
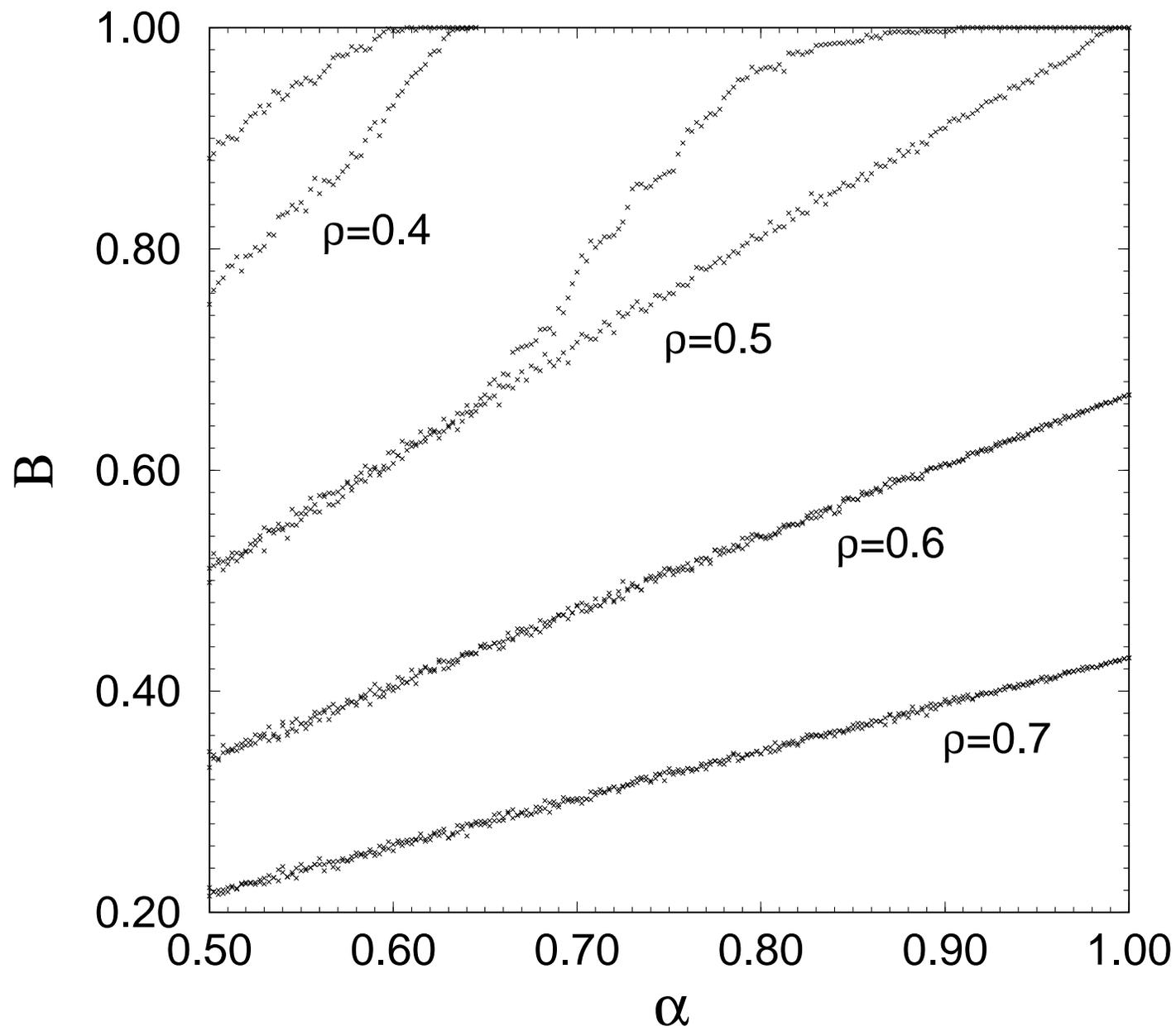
\end{document}